\documentclass{IEEEtran}
\ifCLASSINFOpdf
  \usepackage[pdftex]{graphicx}
  \graphicspath{{figures/}}
  \DeclareGraphicsExtensions{.pdf,.jpeg,.png}
\else
\fi

\usepackage{lipsum}
\usepackage{graphicx}
\ifCLASSOPTIONcompsoc
    \usepackage[caption=false, font=normalsize, labelfont=sf, textfont=sf]{subfig}
\else
\usepackage[caption=false, font=footnotesize]{subfig}
\fi

\usepackage{hyperref}
\hypersetup{
    colorlinks,
    citecolor=black,
    filecolor=black,
    linkcolor=black,
    urlcolor=black
}

\usepackage{cite}
\usepackage{amsmath,amssymb,amsfonts}
\usepackage{algorithmic}
\usepackage{textcomp}
\usepackage{booktabs}
\ifCLASSOPTIONcompsoc
    \usepackage[caption=false, font=normalsize, labelfont=sf, textfont=sf]{subfig}
\else
\usepackage[caption=false, font=footnotesize]{subfig}
\fi

\newcommand{\red}[1]{\textcolor{black}{#1}}
\newcommand{\degree}[1]{$#1^{\circ}$}
\newcommand{\percent}[1]{$#1\%$}
\newcommand{\dist}[1]{$#1mm$}

\begin{document}
\title{Clinical Micro-CT Empowered by Interior Tomography, Robotic Scanning, and Deep Learning}

\author{Mengzhou~Li$^{1+}$, Zheng~Fang$^{1,2+}$, Wenxiang~Cong$^1$, Chuang~Niu$^1$, Weiwen~Wu$^1$, Josef Uher$^{3}$, James Bennett$^4$, Jay T. Rubinstein$^{5}$, and~Ge~Wang$^1$*%
    \\
    $^1$ Department of Biomedical Engineering, Rensselaer Polytechnic, Troy, NY, 12180 USA\\
    $^2$ Department of Instrumental and Electrical Engineering, Xiamen University, Xiamen, 361102 China\\
    $^3$ Radalytica a.s., Prague, 17000 CZE\\
    $^4$ Hawkeye Spectral Imaging, Tucson, AZ, 85701 USA\\
    $^5$ Department of Otolaryngology-HNS, Virginia Merrill Bloedel Hearing Research Center, University of Washington, Seattle, WA, 98122 USA

    \thanks{$^+$ Contributed equally.}
    \thanks{$^*$To whom correspondence should be addressed; E-mail:  wangg6@rpi.edu.}
}

\maketitle
\thispagestyle{empty} %
\pagestyle{empty}

\begin{abstract}
    While micro-CT systems are instrumental in preclinical research, clinical micro-CT imaging has long been desired with cochlear implantation as a primary example. The structural details of the cochlear implant and the temper bone require a significantly higher image resolution than that (about $0.2~mm$) provided by current medical CT scanners. In this paper, we propose a clinical micro-CT (CMCT) system design integrating conventional spiral cone-beam CT, contemporary interior tomography, deep learning techniques, and technologies of micro-focus X-ray source, photon-counting detector (PCD), and robotic arms for ultrahigh-resolution localized tomography of a freely-selected volume of interest (VOI) at a minimized radiation dose level. The whole system consists of a standard CT scanner for a clinical CT exam and VOI specification, and a robotic micro-CT scanner for a local scan \red{of high spatial and spectral resolution at minimized radiation dose.} The prior information from global scan is also fully utilized for background compensation of the local scan data for accurate and stable VOI reconstruction. Our results and analysis show that the proposed hybrid reconstruction algorithm delivers accurate high-resolution local reconstruction, and is insensitive to the misalignment of the isocenter position, initial view angle and scale mismatch in the data/image registration. These findings demonstrate the feasibility of our system design. We envision that deep learning techniques can be leveraged for optimized imaging performance. With high-resolution imaging, high dose efficiency and low system cost synergistically, our proposed CMCT system has great potentials in temporal bone imaging as well as various other clinical applications.
\end{abstract}

\begin{IEEEkeywords}
    Clinical micro-CT, Deep learning, High-resolution imaging, Interior tomography, Photon-counting detector, Robotic arm, Temporal bone imaging, X-ray computed tomography
\end{IEEEkeywords}

\section{Introduction}
\label{sec:introduction}
In the clinical practice of otology and neurotology, medical imaging is critical for evaluation and treatment of many diseases \cite{saadi2020improving}. Magnetic resonance imaging (MRI) is an effective imaging tool with excellent soft tissue contrast and invaluable in detecting  neoplasms with gadolinium contrast agent, but it is not good at evaluating bony structures \cite{rasch1999definition}. Currently, temporal bone CT is the primary method of choice for otological imaging \cite{balachandran2012evaluation}. Chronic otitis media, otosclerosis, temporal bone fracture, congenital aural atresia, cochlear implantation, dehiscent superior semicircular canal, congenital labyrinthine dysplasia, labyrinthine fistula are all disorders or therapies where temporal bone CT is either absolutely necessary or commonly desired adjunct to surgical management, for diagnosis or planning. In cochlear implantation, although psychophysical and physiological measures can be helpful, the ability to localize electrodes and depict their 3D anatomical environment in vivo is particularly instrumental in understanding variations in threshold, uncomfortable level, and channel interaction.

CT is widely used to image a variety of middle and inner ear pathologies but it is limited by suboptimal image resolution \cite{phillips2012interactive}. While most CT scanners have highest spatial resolution of about $0.3~mm$, the latest Cannon CT scanner resolves details down to about $0.2~mm$ \cite{urikura2019overranging}. This level of image resolution is still insufficient. For example, it is common that a precise diagnosis is unavailable until direct assessment during otologic surgery allows visual inspection of the ossicular chain. Also, much-improved image resolution of pre- and post-operative inner ear imaging is required for detailed analysis of cochlear morphometry and its relationship to an implanted electrode array \cite{kjer2018patient}. In these and many other research and clinical applications, there are critical and immediate needs for ultra-fine spatial resolution without compromising other image quality indexes at a minimized radiation dose level. A breakthrough in CT image resolution may greatly benefit diagnosis and treatment in general, and otology and neurotology in particular, such as for rational design and implantation of prosthetic devices.

In contrast to the classic CT system, the robotic-arm-based X-ray imaging system allows a great flexibility of scanning. It supports different scanning trajectories optimized for diverse tasks focusing on various organs and locations. Siemens developed a robotic X-ray system named Multitom Rax. It enables a variety of clinical examinations. Three-dimensional images acquired by Multitom Rax improve diagnostic and planning performance compared to those that do not reflect a natural weight bearing condition. Researchers with Johns Hopkins University developed a mathematical framework for the design of scanning trajectories optimal to a particular task with cone-beam CT \cite{capostagno2019task}. FirstImaging also develop robotic image systems with excellent image quality.

The X-ray photon-counting detector (PCD) is an enabling technology on the horizon for high-resolution (HR) and low-noise imaging, which promises to add a spectral dimension to raw data and boost the CT performance \cite{willemink2018photon}.
Different from the energy-integrating detector (EID), PCD works in a pulse-counting mode and directly converts individual X-ray photons into the corresponding charge signals which are then sorted into different energy bins based on the pulse heights. Thus, the intensity and wavelength information of incoming photons are simultaneously obtained. PCDs principally have no electronic noise and provide significantly small effective pixel size; e.g., around $0.11~mm\times 0.11~mm$. In contrast, EIDs suffer from the dark currents and readout noise with the element size of typically about $1~mm\times1~mm$ due to the point spread function from the propagation of the fluorescence in the scintillator from the interaction position to the visible light sensor. In addition, EIDs put more weights on high-energy photons than low-energy photons but high-energy photons usually get attenuated less than low-energy photons, leading to a reduced image contrast. On the other hand, with PCDs optimal weights can be implemented on polychromatic photons for improved contrast and dose efficiency. More importantly, the energy discrimination ability of PCDs helps reduce beam hardening and metal artifacts, and enables K-edge imaging and material decomposition.

In parallel to the development of X-ray detectors, innovations of micro-focus X-ray tubes are also important. The NanoX tube is an example \cite{hori2018image}, and consists of electron emitting and receiving constructs. The receiving part is mainly an anode with a photoconductor. The emission part includes a backplate, a substrate, a cathode, a gate electrode, and an array of field emission electron sources.
A microstructured array anode target (MAAT) X-ray source is another example, which was recently designed \cite{zan2019design} to offer a significantly higher flux than an ordinary X-ray source in the phase contrast imaging applications.
The key parameters were optimized in the range of 40 keV to 130 keV. This type of technologies could be combined into a micro-focus X-ray tube for temporal bone CT imaging; for example, with a focal spot size of about $0.1~mm$ or less to match the PCD element size.

We envision a robotic-arm-based design of a clinical micro-CT (CMCT) scanner to integrate a micro-focus tube, a PCD, interior tomography, and deep learning into a novel CMCT device that can be abutted to a clinical CT scanner or used separately. The proposed whole CMCT workflow starts with a volumetric scan of a patient, and then proceeds to a subsequent image analysis session by an expert or a smart software analyzer. After that, the patient is smoothly translated into the robotic scanning space for an interior photon-counting micro-CT scan, in registration to the previously performed global scan. Based on interior tomography theory, the interior scan can target only a region of interest (ROI) with a small detector panel and a customized scanning trajectory so that the image quality can be optimized. To facilitate accurate registration between the global scan data/images and local counterparts, an optical 3D surface scanner \red{is used at the start of the local scan and it continues to monitor the head movement of the patient~\cite{li2010experimental} during the local scan for motion compensation}. This design integrates all relevant cutting-edge hardware and software elements, and systematically upgrade our earliest CMCT design published in 2005~\cite{wang2005design} \red{as well as other follow-up designs for similar purposes~\cite{patel2008rotational,chen2009dual,kolditz2010volume,kolditz2012volume,jain2015region,reshef2017dual}}.

\red{Compared to the above-mentioned prior designs, our approach aims at a much higher resolution $~50~\mu m$ with a high-power micro-focus source and a state-of-the-art PCD with fine detector elements for electron-noise-free spectral imaging. Furthermore, the robotic-arms adds a great mechanical flexibility compared to the C-arm gantry and enables a moving VOI mask~\cite{chen2009dual}.}

As far as the image reconstruction is concerned, the current mainstream of interior tomography methods focuses on improving reconstruction quality from truncated projections with no or little extra information~\cite{ye2007general, yu2009compressed, sharma2013scout, jin2014dual, kudo2019advanced,zhu2019interior}. Nevertheless, in our setting a global scan is available, which is clinically natural \red{and can be utilized for robust reconstruction at least for the initial imaging task. Several methods that complete the HR projection data with a low-resolution global scan have been developed for truncation-artifacts-free reconstruction~\cite{azevedo1995region,chun2004x, wang2005design,patel2008rotational,chen2009dual,kolditz2010volume}. Especially, \cite{maass2011new} is computationally-efficient without forward projection operations that directly fuses global and local reconstructions, but it is restricted to a certain type of scanning geometry and inappropriate for our application. Instead of completing the HR truncated projections to cover a global FOV}, we propose to directly perform a background compensation on the HR local/interior scan with the global scan. The main benefit is that the subsequent interior reconstruction only involves a small portion of the sinogram, requiring much less memory space and computational time. This is critical since the projection resolution will be enhanced up to ~10 folds (from ~$1~mm$ to ~$0.11~mm$), hence the amount of interpolated 2D projection data will be increased ~100 times compared to that of the global scan, and handling such big data for reconstruction can be very time-consuming or even prohibitive. \red{Furthermore, for quality enhancement at dose reduction deep learning techniques can be involved in multiple reconstruction stages; e.g., projection deblurring, image denoising and super-resolution, beam hardening correction and material decomposition.}

\section{System Design}

\subsection{System Description}
\begin{figure}[htbp]
    \centering
    \includegraphics[width=0.8\linewidth]{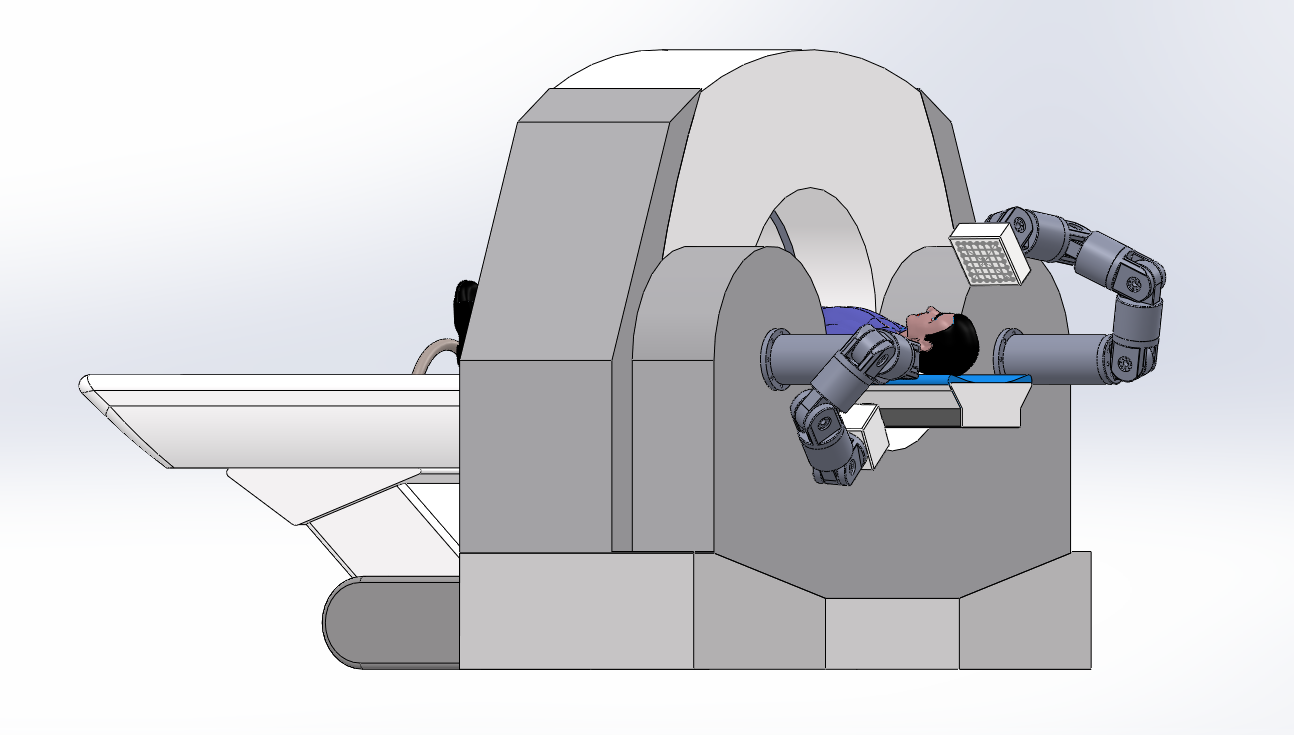}
    \hfill
    \caption{Schematic diagram of the proposed clinical micro-CT (CMCT) system.}
    \label{fig:Schematic}
\end{figure}

The proposed CMCT system consist of a standard medical CT scanner or a novel clincal CT system (such as a NanoX system) and a robotic micro-CT scanner, as shown in Fig.~\ref{fig:Schematic}. The CT scanner performs a global scan. A standard CT algorithm can be used to reconstruct head slices, which show the inner ear region of interest. Then, the patient table transports the patient into the robotic micro-CT system.

\begin{figure}[htbp]
    \centering
    \includegraphics[width=.8\linewidth]{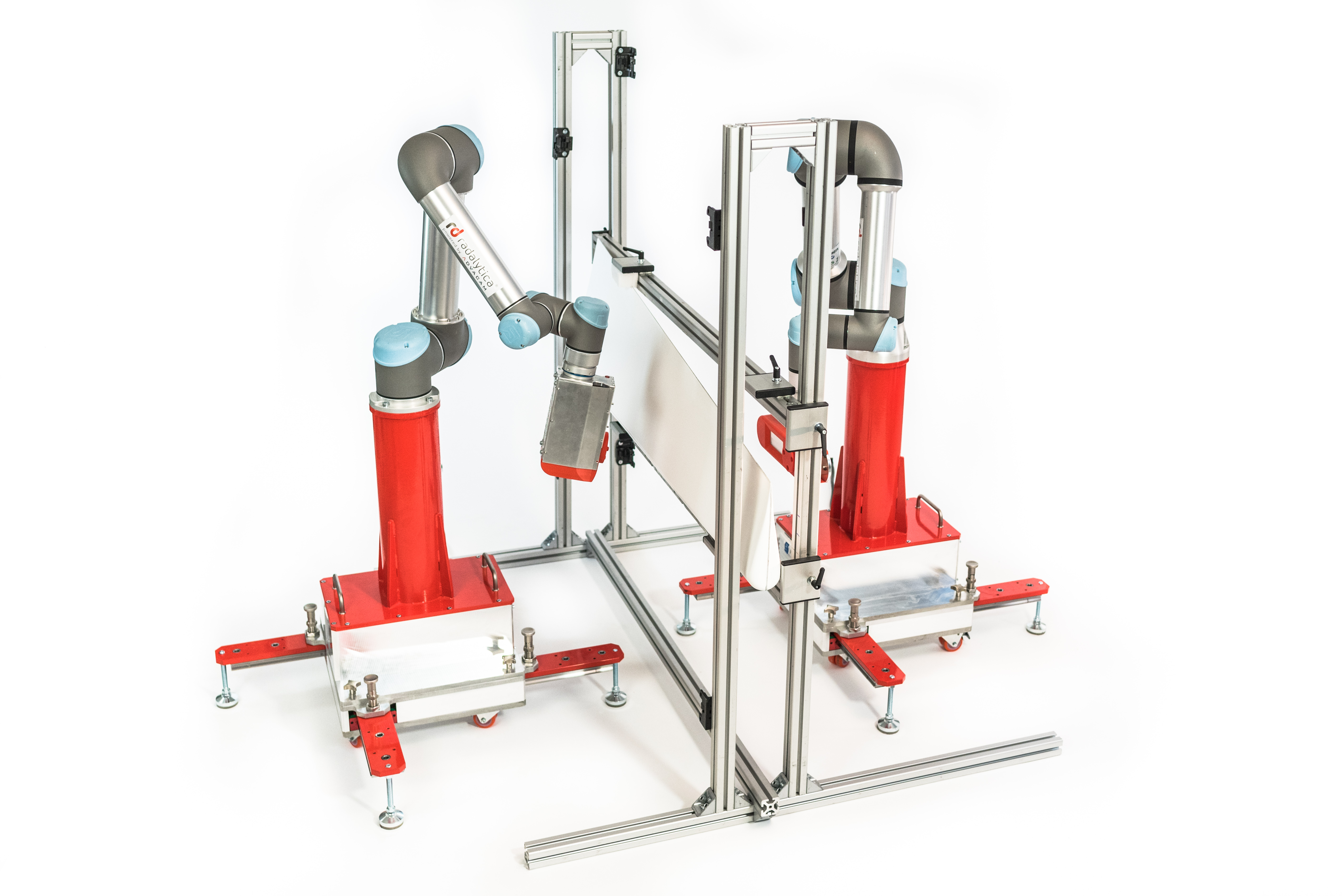}
    \hfill
    \caption{Robotic cone-beam micro-CT system.}
    \label{fig:System}
\end{figure}

This robotic micro-CT system uses two robotic arms holding the X-ray source and the X-ray detector respectively, as shown in Fig.~\ref{fig:System}. The micro-focus tube is a key component for high spatial resolution imaging. The flat-panel PCD is suitable for material decomposition and tissue characterization. The robotic arms can perform a scan along an arbitrary trajectory such as a circular or spiral trajectory. A high-performance computer is in overall control. It sends instructions to the robotic arms, X-ray source and detector. It also acquires raw data from the detector. Each robotic arm needs one control box, which receives commands from the computer and drives the servo system.

\subsection{Design Principles}
\noindent\textbf{Coordinated Robots} The lightweight, highly flexible, and collaborative UR5e and UR16e industrial robot arms with payloads of up to 5 kg and 16 kg respectively are suitable for our CMCT system \cite{subramaniam2020design}. The working radius of UR5e is $850~mm$ with the power consumption of 250 W, while the working radius of UR16e is $900~mm$ with the power consumption of 585 W. The communication between the control computer and the robotic arm control box is through an Ethernet cable using TCP/IP 1000BASE-T protocol. The 6-axis robotic arms have the high position repeatability of $30~\mu m$. Some models can be used to achieve even higher accuracy. The existing Radalytica robotic imaging platform can be modified for CMCT, pushing the limits of the robotic imaging considerably further. The robotic-arm cone-beam micro-CT sub-system can be made to achieve a spatial resolution of ~$50~\mu m$ or higher to meet the temporal bone imaging requirements \cite{li2019novel}. The power supply can be from 100 to 240 VAC at 47 to 440 Hz.

\noindent\textbf{X-ray Source} The Hamamatsu L9181-02 tube is selected as the x-ray source for CMCT. It is an RS-232 controlled 40-130 kV x-ray source with a focal spot size adjustable among 8, 20, and 40 $\mu m$. Its maximum tube power output is 39 W. It combines an X-ray tube high-voltage power supply and a controller in a compact package powered by a 24 VDC, 5A source. The anode target material is Tungsten. The tube window is made of Beryllium of $0.2~mm$ thickness. The effective cone beam angle is ~$45^\circ$. Its weight is 10.5 kg, manageable with the larger robot arm.

\noindent\textbf{Photon-counting Detector} The photon-counting detector ADVACAM WIDEPIX-5x5 of $1280\times1280$ pixels fits the need for human temporal bone micro-CT. Its continuous sensitive surface is supported by an array of $5\times5$ detector tiles. Each tile consists of a single Timepix hybrid detector ($256\times256$ pixels) with an edgeless CdTe sensor. Each pixel counts the number of x-ray photons, allowing a large dynamic range. The Timepix technology also allows for the use of multiple energy thresholds. The intrinsic spatial resolution is defined by the detector pitch of 55 $\mu m$. That is, the imaging sensor covers a $7~cm\times7~cm$ area. The detector weighs 3.3 kg and can be easily carried by the smaller robotic arm. If we use the detector in a $2\times2$ binning mode, image resolution is ~50 $\mu m$, and the diameter of the voulume of interest (VOI) is ~3.5 $cm$, assuming a magnification factor 2. This should be sufficient to cover the human inner and middle ear. In addition to the Timepix detector, we can also customize the Medipix3 detector tiles, whose pixels have two integrated 12-bit digital counters and two energy discrimination thresholds. If we use the detector in a $2\times2$ binning mode, there are 8 spectral bins for data collection in a single scan.

\noindent\textbf{Radiation Dose} Radiation dose is mainly determined by the tube voltage, current and exposure period. With the use of the PCD, there is no electronic noise when recording projection data but Poison noise cannot be avoided. In the interior scanning mode, the X-ray source only radiates about 1/10 of the diameter of the field of view but improvement in image resolution by four times (roughly, from 200 $\mu m$ to 50 $\mu m$) would increase radiation dose significantly (two orders of magnitude) \cite{wang2005design}. Thanks to the latest advancement in deep learning based low-dose CT imaging techniques ~\cite{hendriksen2020noise2inverse}, we can reduce radiation dose by an order of magnitude. All these factors coupled together, on ballpark we should be able to maintain the current head CT dose for an interior micro-CT scan to achieve about 50 $\mu m$ resolution.

\begin{figure}[htbp]
    \centering
    \includegraphics[width=0.8\linewidth]{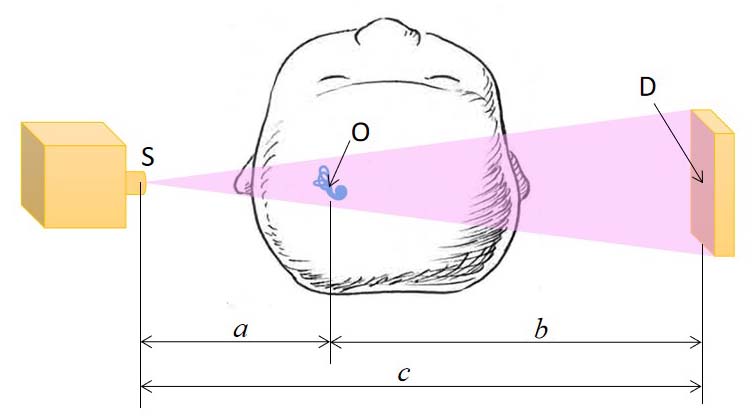}
    \hfill
    \caption{Geometry of a robotic imaging chain.}
    \label{fig:Geometory}
\end{figure}

\noindent\textbf{Spatial Resolution} We define the center of the focal spot as S, the center of the detector as D, and the rotation center as O which is the center of a VOI. These three points should always be kept in a straight line. In Fig.~\ref{fig:Geometory}, $a$ is the source to VOI distance, $b$ is the VOI to detector distance, and $c$ is the source to detector distance. The geometric magnification factors $M$ and $M^\prime$ are $c/a$ and $c/b$ for the focal spot and the detector aperture respectively \cite{feser2008sub}. The spatial resolution $r$ of the imaging system can be approximated as a convolution of the detector size $d$ and the focal spot size $x$ respectively scaled by $M$ and $M'$:

\begin{equation}\label{eq:total_resolutions}
    r = \sqrt{(\frac{x}{M})^2 + (\frac{d}{M'})^2}
\end{equation}

In our initial CMCT system design, $x$ is no more than 40 $\mu m$, and d is equal to 110 $\mu m$ (due to $2 \times 2$ binning). The source to VOI distance can be from 150 to $250~mm$, and the VOI to detector distance is in the same range. Consequently, the magnification factor can be adjusted from 1.60 to 2.67 with the imaging field of view from 26 to 44 $mm$ in diameter. According to the above equations, the system spatial resolution can be made ~50 $\mu m$.

\begin{figure}[htbp]
    \centering
    \includegraphics[width=1\linewidth]{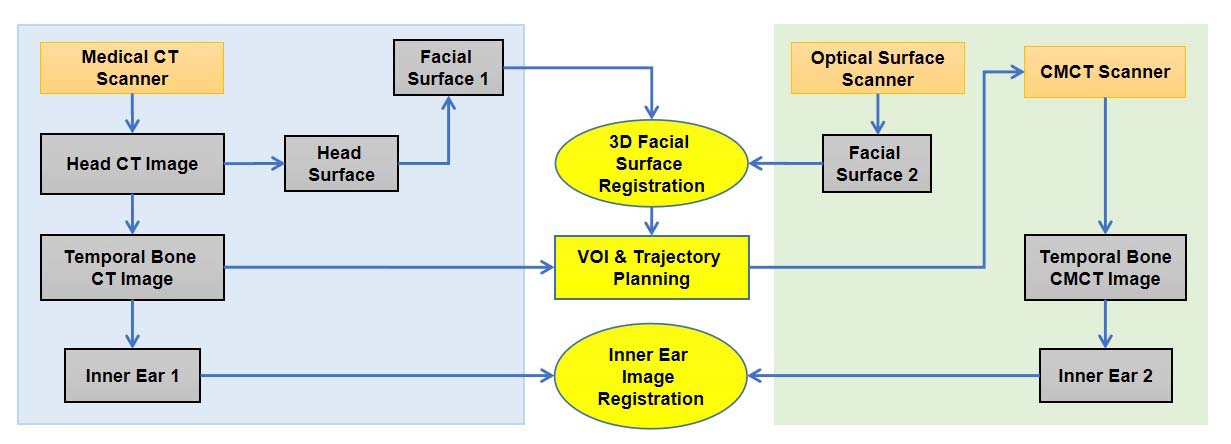}
    \hfill
    \caption{CMCT image registration integrating standard temporal bone CT and our robotic micro-CT imaging.}
    \label{fig:seg}
\end{figure}

\noindent\textbf{Geometric Alignment} In the CMCT process, global and local projection data are obtained in different scanning geometries. An optic 3D surface scanner is preferably used for image registration, as shown in Fig.~\ref{fig:seg}.  The clinical CT scan obtains a 3D image of a patient who may wear landmarks such as a firmly-attached helmet. A boundary detection algorithm extracts the facial/helmet surface and key points for mesh generation as a basis for image registration \cite{bakirman2017comparison}. The fact is utilized that the inner ear and the face surface are in a rigid relation. As an example of 3D surface scanners, the laser scanner Micro-Epsilon LLT2910-100 is a high-quality profile sensor. The height range may be set from 125 to 390 $mm$,  and the width range can be fixed to $143.5~mm$,  with spatial resolution of 12 $\mu m$ at a profiling frequency of 300 Hz. The compact size of $96~mm \times 85~mm \times 33~mm$ and low weight of 380 grams are ideal for static, dynamic and robotic applications. The wavelength of the semiconductor laser is switchable between $658~nm$ (red) and $405~nm$ (blue). The measurement data can be output via an Ethernet UDP, Modbus TCP, or serial communication protocol. The power supply is 24 VDC 500 mA. The optical scanner and the two robotic arms are mounted on the same pedestal so that they share the same coordinate system. The optically scanned patient head surface needs to be registered with the clinical CT originated head surface. Then, the coordinate conversion is carried out to delineate a VOI for a robotic micro-CT scan. The temporal bone data and images by medical CT and micro-CT also need to be registered and fused to achieve the best imaging performance.

\section{Imaging Algorithm}\label{sec:reconSTG}
It is well known that the CT interior problem is not uniquely solvable in an unconstrained space \cite{natterer2001mathematics}. By introducing additional prior knowledge on the image to be reconstructed; e.g., an interior sub-region with known attenuation values or a piece-wise constant model of underlying images, the image reconstruction from local projection profiles that are truncated on both sides becomes uniquely solvable. However, those assumptions often do not exactly hold in practical cases, which potentially results in shifting and cupping in reconstructed attenuation values. To address this problem, we propose to use a low-resolution (LR) global CT scan to estimate the background attenuation in the sinogram of the HR local scan that involves the surrounding volume of the VOI, and obtain an accurate HR local reconstruction of the VOI.

\begin{figure}[htbp]
    \centering
    \subfloat[global scan\label{fig:1a}]{%
        \includegraphics[width=0.8\linewidth]{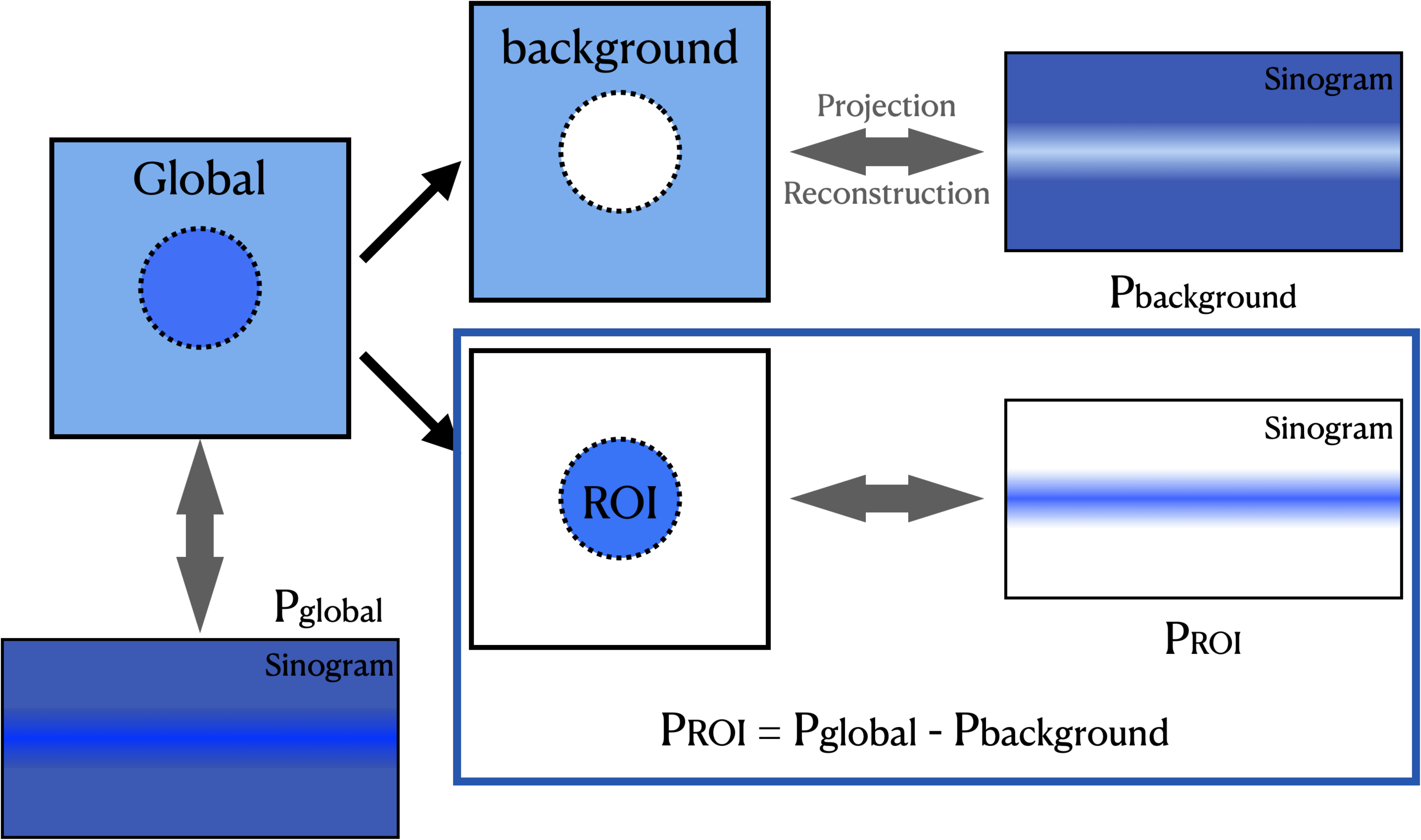}}
    \hfill
    \subfloat[local scan\label{fig:1b}]{%
        \includegraphics[width=0.8\linewidth]{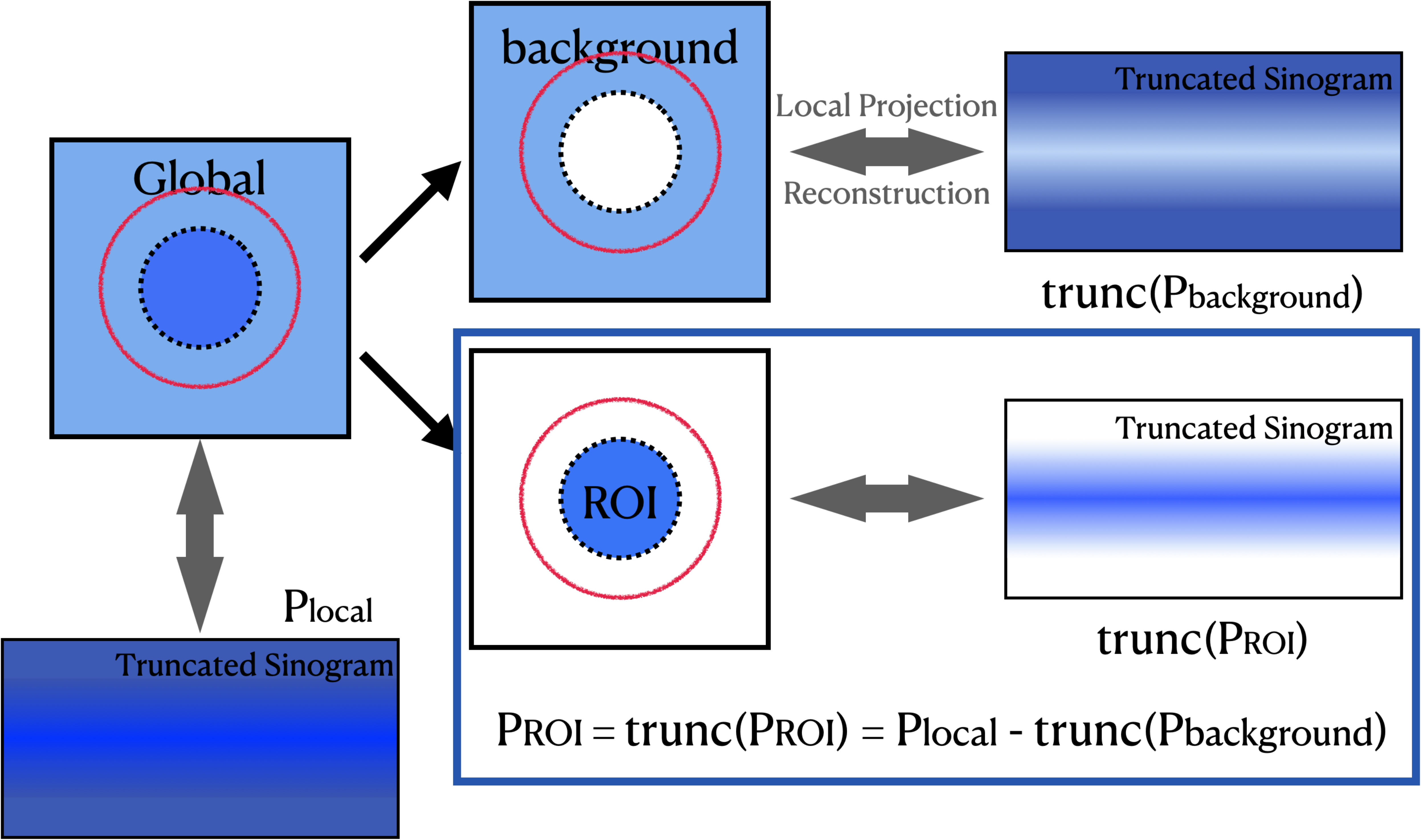}}
    \caption{Illustration of the proposed ROI reconstruction through background compensation.}
    \label{fig:basicPrinciple}
\end{figure}

The general idea of background compensation is illustrated in Fig.~\ref{fig:basicPrinciple} in a 2D case without loss of generality. The transformation between the underlying/reconstructed image and its projection data is {linear} and invertible, and we can partition an underlying image into the two parts, the region of interest and the background. Given the sinogram of the background $P_{background}$ and the global sinogram $P_global$, we can easily obtain the pure sinogram of the region of the interest (ROI) $P_{ROI}$ as follows:
\begin{equation}\label{eq:bgcrt}
    P_{ROI} = P_{global} - P_{background}.
\end{equation}
The above relationship becomes nontrivial with laterally truncated projection data. Let $trunc(\cdot)$ denote the truncation operation, and Eq.~\ref{eq:bgcrt} becomes
\begin{equation}\label{eq:bgcrtTrunc}
    trunc(P_{ROI}) = trunc(P_{global}) - trunc(P_{background}),
\end{equation}
where $trunc(P_{global})$ stands for a local scan $P_{local}$. By intentionally letting the local scan cover the ROI, the truncated parts of $P_{ROI}$ are all zeros, and we have
\begin{equation}
    P_{ROI} = P_{local} - trunc(P_{background}).
\end{equation}
This equation suggests that the ROI within a local scan can be accurately reconstructed from the laterally truncated scan after background subtraction. Clearly, this compensation will improve the stability of interior tomography.

For CMCT, a VOI can be accurately reconstructed at high resolution, given the HR local projection dataset $P_{local}^{HR}$ and an appropriate background estimation. Suppose that we have a prior LR CT scan of the object $P_{global}^{LR}$, the LR background estimation $P_{background}^{LR}$ can be easily obtained as shown in Fig.~\ref{fig:basicPrinciple}(a), which can be used to approximate the HR background closely via interpolation. This is based on the assumption that residual high-frequency background estimation errors will mostly cancel out during the integration which makes the LR estimation sufficiently accurate for our purpose.

\subsection{CMCT Reconstruction}
In clinical applications, a standard (relatively low resolution) global CT scan is first performed. Those with pathological potentials can be further examined with a local micro-CT scan, which provides a HR local/interior reconstruction of a VOI. With interior tomography, a minimized additional dose will be involved. The prior information obtained through the global CT scan can be utilized to help the interior image reconstruction (at ultrahigh resolution).

The whole scanning procedure is summarized as follows:
\begin{enumerate}
    \item Perform a global head CT scan $P_{global}$ with a scanning geometry $G_{global}$;
    \item Determine a VOI and then plan a scanning geometry $G_{local}$ for a local micro-CT scan;
    \item Scan the patient optically with a surface scanner to generate the surface model $S_{local}$ for data/image registration between the global and local scans at the start of the local scan;
    \item Perform the micro-CT scan $P_{local}$ following $G_{local}$ (see the registration step below). \red{During the local scan, the optical scanner continuously scans the patient head} (preferably with optical markers) for tracing unconscious head movement.
\end{enumerate}

Then, the VOI reconstruction mainly consists of the following three steps: data/image registration, background compensation, and image reconstruction.
\red{For brevity, in the following we assume that the head movement effects have already been compensated for in the local projections $P_{local}$ with the feedback from the optical scanner using advanced correction techniques, like, locally linear embedding motion correction~\cite{chen2016x,chen2018general}}.
\begin{enumerate}
    \item \textbf{Registration}: Find the relative geometry between $G_{global}$ and $G_{local}$ in reference to the facial surface model $S_{local}$.
          \begin{enumerate}
              \item Reconstruct the global volume $V_{global}$ from $P_{global}$;
              \item Render the surface model $S_{global}$ from the global reconstruction $V_{global}$;
              \item Register the two surface models to align the orientation and position of $S_{global}$ with $S_{local}$. \red{The registration result is used to guide the micro-CT scan}\label{step};
              \item From $P_{local}$, directly reconstruct a volume of interest $V_{local}$ which contains fine structures but may be subject to distorted attenuation values;
              \item \red{Refine the registration parameters (obtained in Step~\ref{step})} in reference to the registration between $V_{global}$ and $V_{local}$.
          \end{enumerate}
    \item \textbf{Compensation}: Correct the attenuation offsets in $P_{local}$ to form a pure sinogram of the VOI $P_{VOI}$.
          \begin{enumerate}
              \item Set the attenuation values inside VOI to zero in the aligned global reconstruction to form the background volume $V_{background}^{align}$;
              \item Digitally reproject the background volume $V_{background}^{align}$ following the geometry $G_{local}$ to form the LR background estimation $P_{background}^{LR}$;
              \item Interpolate $P_{background}^{LR}$ to the same resolution as the local HR projection of CMCT $P_{local}$, and obtain $P_{background}^{HR}$;
              \item Truncate $P_{background}^{HR}$ to the same size as $P_{local}$, and obtain $trunc(P_{background}^{HR})$;
              \item Correct $P_{local}$ with the estimated attenuation background as $P_{local} - trunc(P_{background}^{HR})$ to form the pure sinogram of the VOI $P_{VOI}$;
          \end{enumerate}
    \item \textbf{Reconstruction}: Reconstruct the VOI from $P_{VOI}$ with geometry $G_{local}$ using a cone-beam reconstruction algorithm, and preferably one developed in the deep learning framework.
\end{enumerate}

\section{Results}
\subsection{CMCT Accuracy and Resolution}
\begin{figure*}[htbp]
    \centering
    \includegraphics[width=0.7\linewidth]{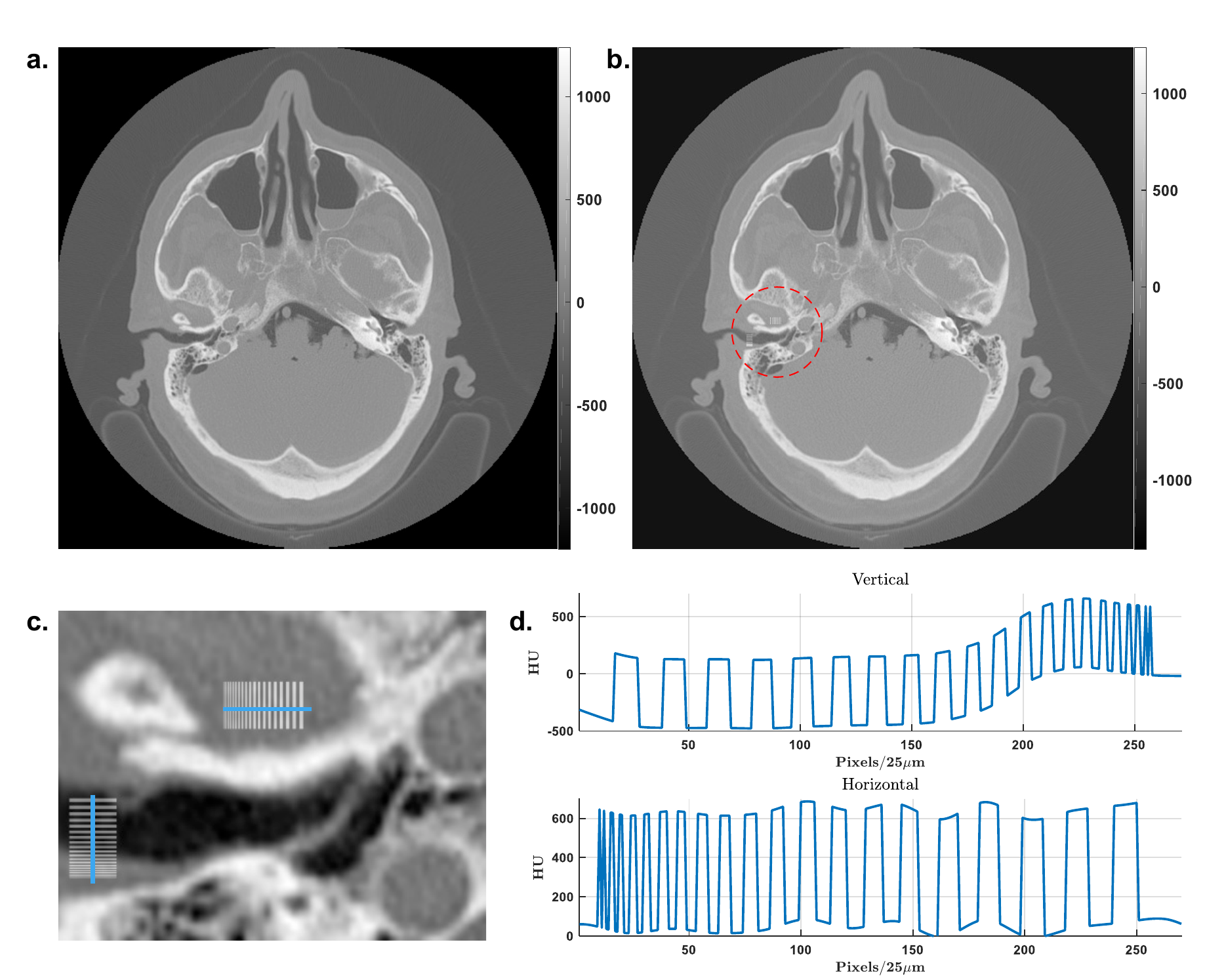}
    \caption{Head phantom with resolution patterns. (a) The original head image, (b) two resolution bar patterns placed along horizontal and vertical directions respectively with an amplitude of 600 HU in the inner ear region, (c) an enlarged region around the patterns, and (d) the profiles of the resolution patterns.}
    \label{fig:phantom}
\end{figure*}

\begin{figure*}[htbp]
    \centering
    \includegraphics[width=0.9\linewidth]{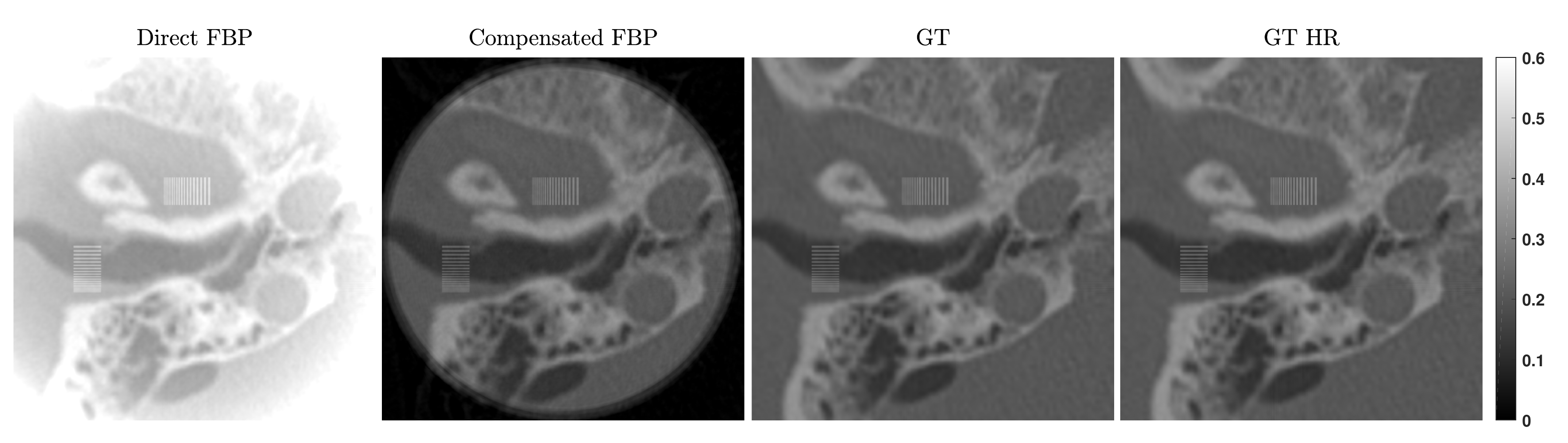}
    \caption{Reconstructed ROI in the same display window, including the direct reconstruction with FBP from local data, FBP reconstruction after background compensation, the ground truth reconstructed from the HR global scan with $0.11mm$ pixel size, and the HR ground truth from the HR global scan with $0.04mm$ pixel size. The attenuation value is in $cm^{-1}$.}
    \label{fig:recons}
\end{figure*}

A simulation study was performed to demonstrate the feasibility of our proposed VOI reconstruction through background compensation. When the PCD is used, projections are collected in a number of energy bins. For inner ear imaging, we may initially focus on all the counts in a wide energy window to study the reconstruction performance; i.e., 40keV to 110keV (120 kVp source), for the following considerations: (1) to avoid the blurring from the X-ray fluorescence in the CdTe crystal; (2) to collect most photons that have penetrated through the head; and (3) to reduce the influence from the pile-up effects. One head CT image containing the inner ear structures from the Visible Human Project \cite{ackerman1998visible} was used as a realistic image phantom, as shown in Fig.~\ref{fig:phantom}(a). The original phantom matrix is of $512 \times 512$ pixels with pixel size of $0.4981~mm$. In our study, the phantom image was first interpolated to $10,240 \times 10,240$ pixels with a pixel size of $0.025~mm$ to generate the HR phantom. To be mentioned, the bicubic interpolation method was used to generate pixel values. Then, two resolution bar patterns in horizontal and vertical orientations were embedded in the inner ear region as shown in Fig.~\ref{fig:phantom}(b) to (d). The amplitude of the added patterns is 600 HU, and the radius of the ROI is $23~mm$ as marked with a red circle in Fig.~\ref{fig:phantom}(b). A global CT scan was performed with $1.024~mm$ detector pixel size, and the HR local CT scan covered a region of radius $35.05~mm$ with a detector pixel size $0.11~mm$. Both scans had the same system magnification factor of 2 and tube voltage of 140kVp, and were simulated with an industrial CT simulator CatSim \cite{de2007catsim}. Two additional HR global scans were also performed with fine detection pixels of $0.11~mm$ and $0.04~mm$ to produce the reference reconstructions as the ground truth (GT) of the attenuation value and image resolution, respectively.
In Fig.~\ref{fig:recons}, large deviations of attenuation values from the GT and strong cupping effects are observed for the direct local scan reconstruction with filtered back projection (FBP). In contrast, the reconstruction from the background compensated projections demonstrates accurate attenuation values. The negligible artifacts around the ROI boundary may come from minor mismatches between the background estimation. Interestingly, for the direct reconstruction, in spite of the attenuation value shifting, the fine details are still clearly discernible except for the distorted ROI boundary, which might be sufficient in those applications that only need structural features.

\begin{figure}[htbp]
    \centering
    \includegraphics[width=1.\linewidth]{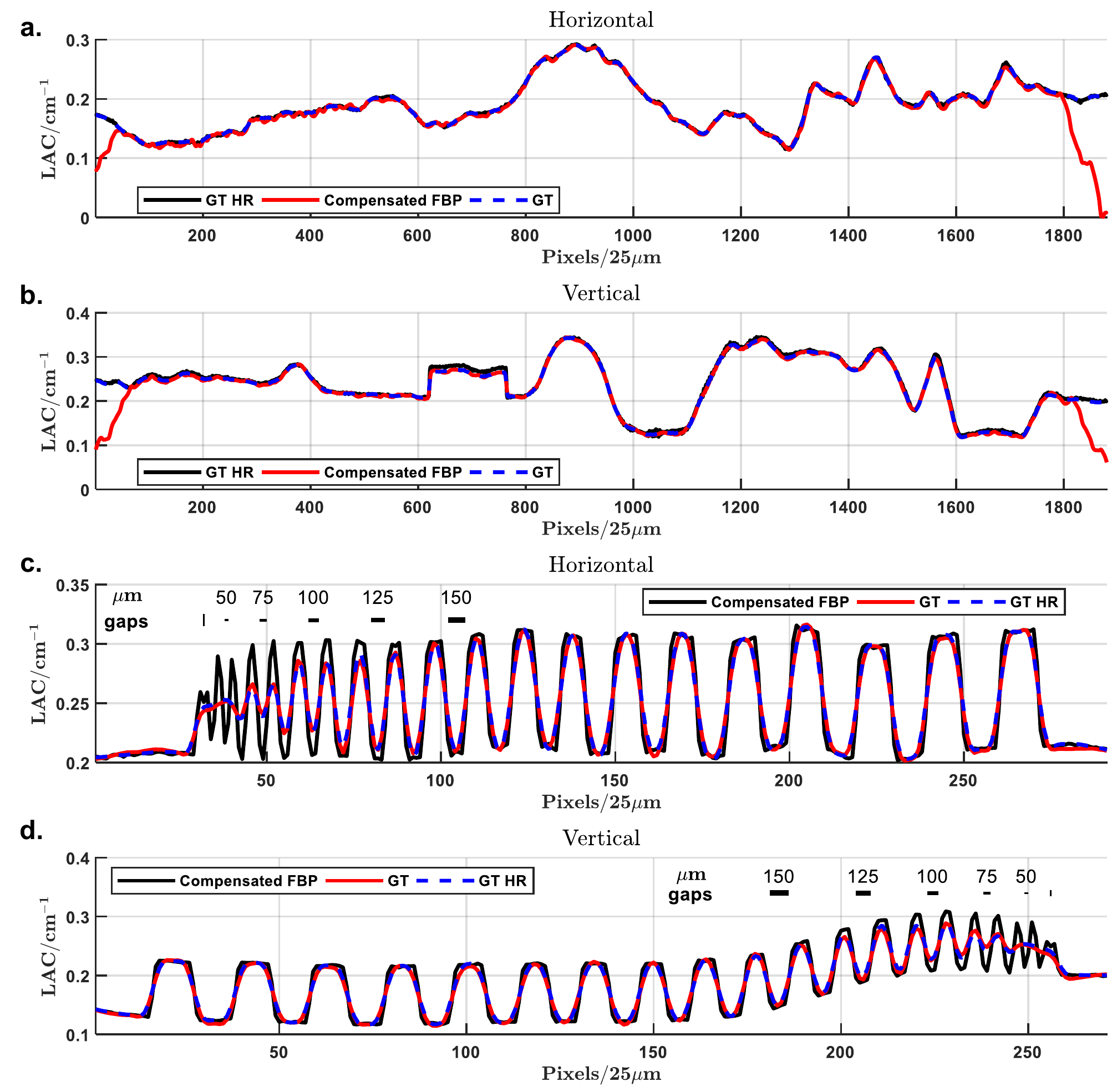}
    \caption{Attenuation profiles of the local reconstructions after background compensation relative to the ground truth. (a) and (b) The profiles along the horizontal and vertical midlines for attenuation value comparison; (c) and (d) the profiles through the horizontal and vertical resolution bar patterns for image resolution assessment.}
    \label{fig:profiles}
\end{figure}

To evaluate the potential influence of the proposed method on the attenuation value and image resolution, the profiles along and around the vertical and horizontal midlines of the ROI and through the resolution bar patterns are shown in Fig.~\ref{fig:profiles}. The lines overlap well with the ground truth inside the ROI as demonstrated in Fig.~\ref{fig:profiles}(a) and \ref{fig:profiles}(b), demonstrating the high fidelity of the reconstructed attenuation values. Similarly in Fig. ~\ref{fig:profiles}(c) and (d). The overlapped profiles of the GT and FBP reconstruction show that the proposed local reconstruction method does not compromise image resolution as compared with the global HR scan. Particularly, the $50~\mu m$ gap is not resolved while the $75~\mu m$ gap is well resolved, which indicates the resolution of the local scan protocol is between $50~\mu m$ and $75~\mu m$. This agrees well with our analysis on the imaging parameters.

\subsection{Mis-registration Effects}\label{sec:Misalign}
While the accuracy and resolution have been demonstrated above, the feasibility of the proposed reconstruction method will be illustrated in this subsection in terms of the robustness to the a potential mis-registration due to imperfect hardware components and their suboptimal coordination; i.e., with respect to mismatches in position, orientation and/or scale. The direct effects of these mismatches on the reconstruction process are an isocenter offset , a falsely tilted initial view angle, and an incorrect magnification factor messing up the interpolation between the local scan and the re-projection through a globally reconstructed image volume for background estimation.

\begin{figure}[htbp]
    \centering
    \includegraphics[width=1.\linewidth]{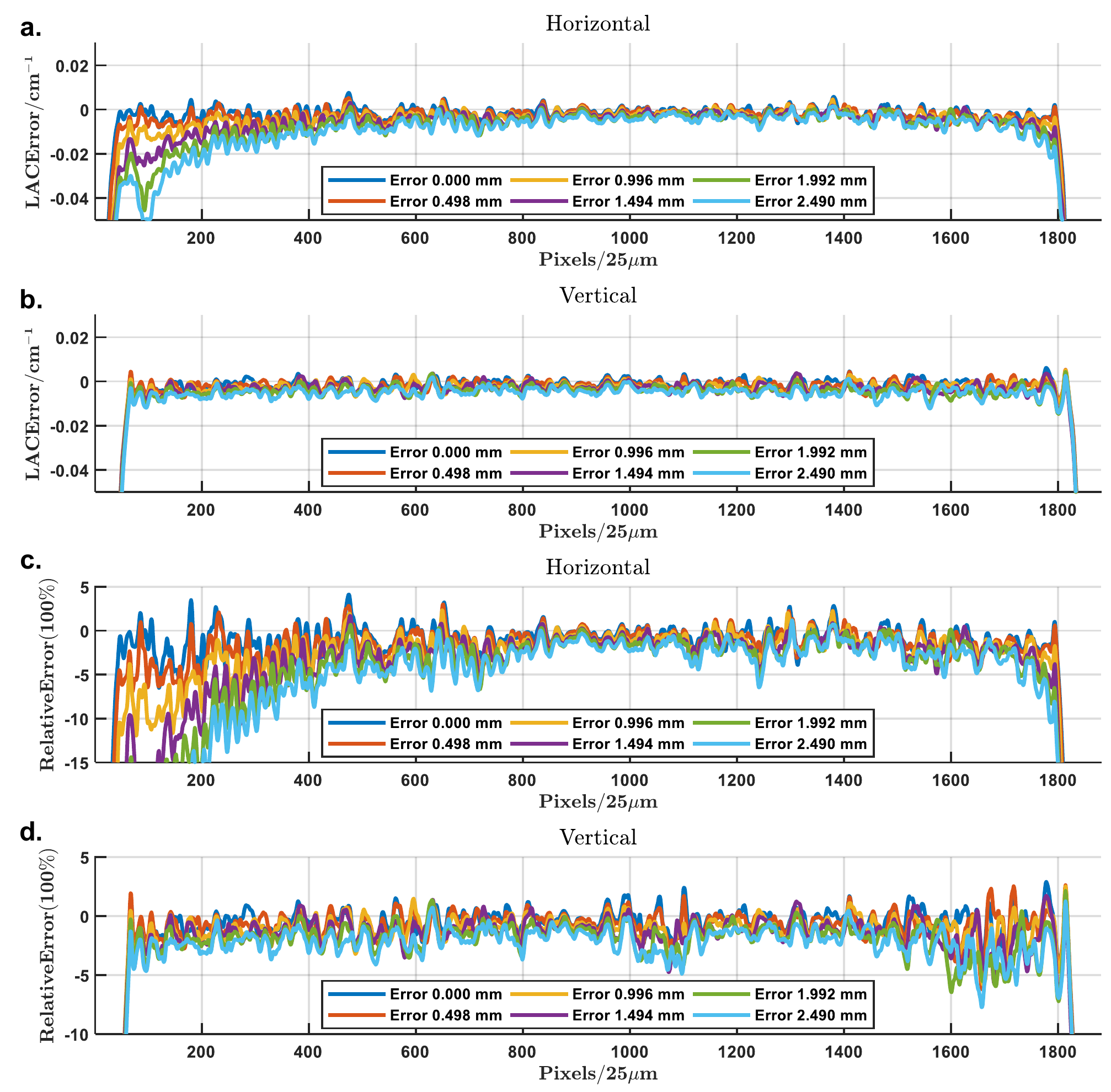}
    \caption{Profiles of the attenuation error in the reconstructions with isocenter misalignment after background compensation. (a) The horizontal and (b) vertical profiles of the attenuation error through the VOI center; (c) and (d) the corresponding relative error profiles. The profiles are associated with isocenter mismatches from 0 to $2.490~mm$ in a step of $0.498~mm$. }
    \label{fig:PrLines}
\end{figure}

First, the mismatches in the isocenter position were set to the range from 0 to $4.483~mm$ with an increment of $0.498~mm$ along the horizontal direction. The corresponding results with misaligned background compensation are in Fig.~\ref{fig:Pimgs}. The absolute and relative error maps were calculated against the GT. Then, the profiles along the horizontal and vertical midlines of the error maps corresponding to 6 selected mismatches are in Fig.~\ref{fig:PrLines}. It is observed that the maximum relative error of the compensated local reconstruction is within $\pm5\%$, which came mainly from the error in the background estimation, while the misalignment along the horizontal direction seems having little effect on the vertical profiles while there are significant drops off the horizontal profiles. Specifically, the maximum relative error remains around $6\%$ when the misalignment is $0.498~mm$, and it becomes close to $10\%$ when the misalignment reaches $0.996~mm$. The above observations suggest the robustness of the compensated local reconstruction with respect to the isocenter misalignment which remains quite accurate even with an up to $0.498~mm$ mismatch.

\begin{figure}[htbp]
    \centering
    \includegraphics[width=1.0\linewidth]{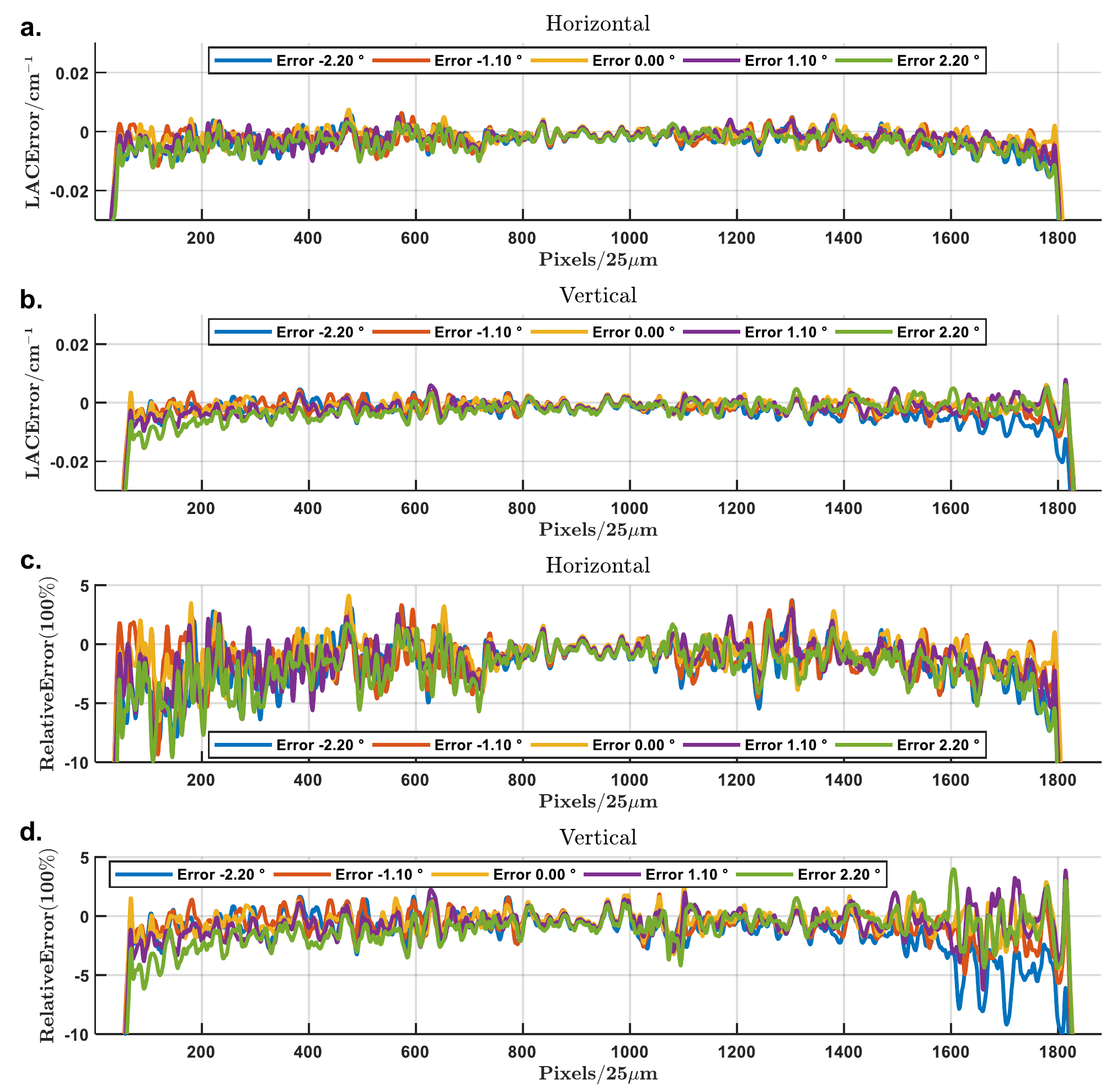}
    \caption{Profiles of the attenuation error in the reconstructions with misaligned initial view angles for background compensation. (a) and (b) The horizontal and vertical profiles through the center of the attenuation error map, (c) and (d) the corresponding relative reconstruction error profiles. The profiles correspond to angular mismatches from $-2.2^{\circ}$ to $2.2^{\circ}$ with an increment of $1.1^{\circ}$.}
    \label{fig:ArLines}
\end{figure}

The initial view angle mismatches were simulated in the range from $-4.39^{\circ}$ to $4.39^{\circ}$ with an increment of $1.098^{\circ}$. The resultant reconstructions are in Fig.~\ref{fig:Aimgs}. The error maps were calculated in the same way as for the positional mismatches, and the through-center profiles of the error maps are in Fig.~\ref{fig:ArLines}. Similarly, the mismatches have stronger influences on the horizontal profiles than on the vertical ones, by the directional asymmetry of the background compensation. The maximum relative error is in the range $[-10\%, 5\%]$ when the angular errors are within $\pm2.2^{\circ}$. Especially, the relative reconstruction errors are mostly contained within $\pm6\%$ if the angular errors are within $\pm1.1^{\circ}$.

\begin{figure}[htbp]
    \centering
    \includegraphics[width=1.0\linewidth]{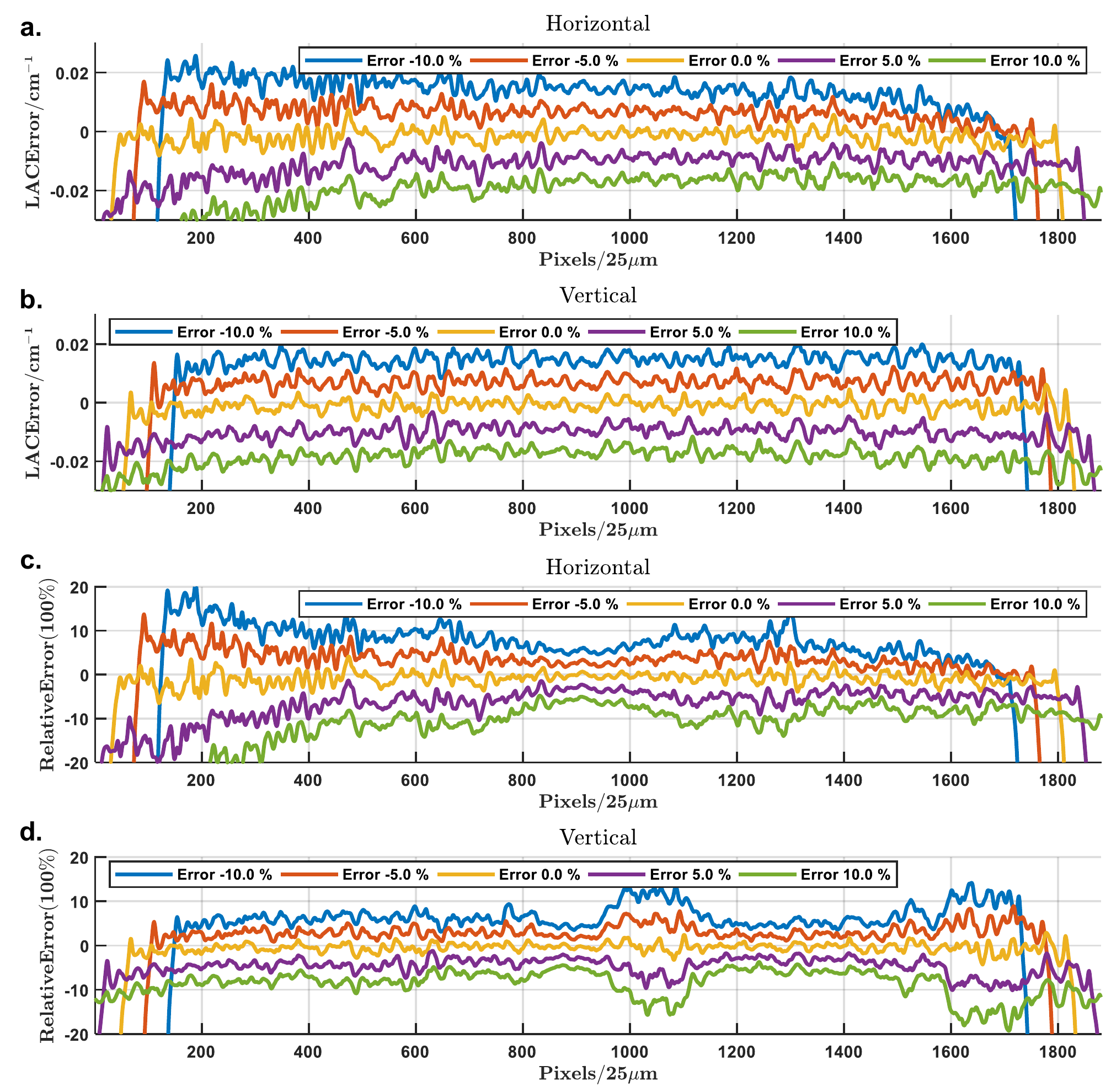}
    \caption{Profiles of the attenuation error in the reconstructions with incorrect magnification factors for background compensation. (a) and (b) The horizontal and vertical profiles through the reconstruction error maps; (c) and (d) the corresponding relative reconstruction error profiles. The profiles correspond to magnification mismatches from $-10\%$ to $10\%$ with an increment of $5\%$.}
    \label{fig:MrLines}
\end{figure}

The magnification errors were also simulated in the range from $-20\%$ to $20\%$ in a step of $5\%$, and the corresponding reconstructions are in Fig.~\ref{fig:Mimgs}. Interestingly, the ``effective'' ROI of reconstruction is determined by the applied magnification factor, with the regions outside the effective ROI having opposite attenuation shifting as compared to that inside the effective ROI. The relative reconstruction error profiles are in Fig.~\ref{fig:MrLines} with the corresponding magnification mismatches from $-10\%$ to $10\%$. Different from the distortions observed in the cases of positional and angular misalignments, which are mainly concentrated around a peripheral region, the magnification error causes a global attenuation shift inside the effective ROI. As shown in Fig.~\ref{fig:MrLines}(a) and (b), the attenuation shift is proportional to the magnification mismatch \red{and seems much more sensitive than other types of misalignments. Fortunately, this global shift can be effectively addressed by a bias correction method introduced below.}

\subsection{Bias Correction}
The global shift, especially observed in the case with magnification errors, can be addressed with bias correction to make the attenuation value of a known region agree with the target value; i.e., to make the air value close to zero. Other known values can be used for the same purpose, such as that obtained from the global reconstruction; for example, we can select a relatively flat region and calculate its mean value as a benchmark.

Quantitative metrics, including SSIM, Peak Signal-to-Noise Ratio (PSNR), MSE and Root Mean-Squared Relative Error(RMSRE), were used to quantify the reconstructions with these misalignments in reference to the ROI in the GT, with and without additional bias correction. The actual radius of the ROI was set to $21~mm$. The evaluation results with respect to isocenter position, initial angle, and magnification errors are summarized in Tables~\ref{table:QM_P}, \ref{table:QM_A} and \ref{table:QM_M} respectively.
The results after bias correction are denoted with the abbreviation `Crt' in the Tables.

\red{The reconstructions are quite robust with respect to the isocenter positional and initial angular errors, and the bias correction method can further improve the accuracy}, as shown in Tables~\ref{table:QM_P} and \ref{table:QM_A}. The attenuation deviation from GT in the reconstruction with aligned background compensation is very small, with MSE $0.454\times 10^{-5}$ and RMSRE only \percent{1.07}. By increasing the position error, the SSIM and PSNR metrics decrease while MSE and RMSRE increase, as expected. The RMSRE value is still below \percent{2.0} when the position error reaches $0.996~mm$, \red{and the tolerance is extended to up to $1.992mm$ after the bias correction}, which demonstrate the robustness of our method. Similarly, in Table~\ref{table:QM_A} the RMSRE remains below \percent{2.0} for all angular errors within \degree{\pm4.39}.

\red{The bias correction substantially improves the magnification-error-affected image reconstruction}, as shown in Table~\ref{table:QM_M}. To be noted, since the effective ROI is scaled with the magnification factor, the intersection of these effective ROIs and the original ROI (radius $21~mm$) was used for evaluation, with the radius being set to $16.8~mm$. The metrics on the reconstruction before the bias correction dramatically change as the magnification error varies, demonstrating a relative high sensitivity. Fortunately, with the bias correction method RMSRE in the case of \percent{-20} magnification error is significantly reduced to \percent{1.67} from \percent{10.43} within the effective ROI, \red{demonstrating a decent robustness}.

An interesting phenomenon is that if we perform the bias correction on the aligned reconstruction, \red{the metric scores drop slightly in the inner region within the radius of $16.8~mm$ in contrast to a tiny boost when evaluated on the whole ROI region (radius $21~mm$)}. This suggests that the reconstruction must already be very accurate in the inner region when the compensation is well aligned. The attenuation estimation from the global reconstruction cannot be perfect due to differences in resolution and existence of artifacts, \red{and the linearity of the model to support our compensation method is just an approximation to the polychromatic X-ray imaging process, which may bring residual errors into the peripheral region of the ROI}. Thus, there is no need to perform bias correction if the system is well calibrated and we are sure about the accuracy of the registration for background compensation. Otherwise, bias correction can be used for better performance.

\begin{table*}[htbp]
    \caption{\label{table:QM_P} Quantitative metrics of the reconstructions with isocenter misalignment for background compensation (r=$21mm$)}
    \renewcommand{\arraystretch}{1.}
    \begin{center}
        \resizebox{0.8\linewidth}{!}{
            \begin{tabular}{lcccccccccc}
                \toprule
                Position Error            & aligned    & \dist{0.498} & \dist{0.996} & \dist{1.494} & \dist{1.992} & \dist{2.490} & \dist{2.989} & \dist{3.487} & \dist{3.985} & \dist{4.483} \\
                \midrule
                SSIM                      & \bf0.9998  & 0.9998       & 0.9996       & 0.9993       & 0.9989       & 0.9984       & 0.9977       & 0.9970       & 0.9961       & 0.9952       \\
                PSNR                      & \bf46.2150 & 44.9155      & 42.3074      & 39.7698      & 37.6195      & 35.8479      & 34.4174      & 33.2073      & 32.1431      & 31.2084      \\
                MSE$\times10^{-4}$        & \bf0.0454  & 0.0613       & 0.1117       & 0.2004       & 0.3288       & 0.4944       & 0.6873       & 0.9081       & 1.1603       & 1.4389       \\
                RMSRE                     & \bf0.0107  & 0.0125       & 0.0171       & 0.0231       & 0.0298       & 0.0367       & 0.0434       & 0.0500       & 0.0566       & 0.0631       \\
                \hline
                SSIM (Crt.)               & \bf0.9998  & 0.9998       & 0.9998       & 0.9997       & 0.9996       & 0.9993       & 0.9991       & 0.9988       & 0.9984       & 0.9980       \\
                PSNR (Crt.)               & \bf46.3106 & 46.2296      & 45.1858      & 43.6330      & 41.8483      & 40.1459      & 38.7410      & 37.5219      & 36.4183      & 35.4465      \\
                MSE$\times10^{-4}$ (Crt.) & \bf0.0444  & 0.0453       & 0.0576       & 0.0823       & 0.1242       & 0.1838       & 0.2540       & 0.3363       & 0.4336       & 0.5423       \\
                RMSRE (Crt.)              & \bf0.0104  & 0.0105       & 0.0120       & 0.0146       & 0.0182       & 0.0225       & 0.0266       & 0.0308       & 0.0350       & 0.0393       \\
                \bottomrule
            \end{tabular}
        }
    \end{center}
\end{table*}

\begin{table*}[htbp]
    \caption{\label{table:QM_A} Quantitative metrics of the reconstructions with initial angular misalignment for background compensation (r=$21~mm$)}
    \renewcommand{\arraystretch}{1.}
    \begin{center}
        \resizebox{0.7\linewidth}{!}{
            \begin{tabular}{lccccccccc}
                \toprule
                Angle Error               & \degree{-4.39} & \degree{-3.29} & \degree{-2.20} & \degree{-1.10} & aligned    & \degree{1.10} & \degree{2.20} & \degree{3.29} & \degree{4.39} \\
                \midrule
                SSIM                      & 0.9991         & 0.9994         & 0.9996         & 0.9998         & \bf0.9998  & 0.9998        & 0.9996        & 0.9994        & 0.9991        \\
                PSNR                      & 38.4776        & 40.0098        & 42.1055        & 44.8752        & \bf46.2150 & 44.3655       & 41.9105       & 39.9452       & 38.4152       \\
                MSE$\times10^{-4}$        & 0.2699         & 0.1896         & 0.1170         & 0.0619         & \bf0.0454  & 0.0696        & 0.1224        & 0.1925        & 0.2738        \\
                RMSRE                     & 0.0273         & 0.0226         & 0.0176         & 0.0127         & \bf0.0107  & 0.0128        & 0.0166        & 0.0206        & 0.0244        \\
                \hline
                SSIM (Crt.)               & 0.9995         & 0.9997         & 0.9998         & 0.9998         & \bf0.9998  & 0.9998        & 0.9998        & 0.9997        & 0.9995        \\
                PSNR (Crt.)               & 41.8476        & 43.3071        & 44.8716        & 46.1801        & \bf46.3105 & 46.1733       & 45.0336       & 43.3808       & 41.7435       \\
                MSE$\times10^{-4}$ (Crt.) & 0.1242         & 0.0887         & 0.0619         & 0.0458         & \bf0.0444  & 0.0459        & 0.0596        & 0.0873        & 0.1272        \\
                RMSRE (Crt.)              & 0.0179         & 0.0149         & 0.0123         & 0.0105         & \bf0.0104  & 0.0108        & 0.0121        & 0.0144        & 0.0172        \\
                \bottomrule
            \end{tabular}
        }
    \end{center}
\end{table*}

\begin{table*}[htbp]
    \caption{\label{table:QM_M} Quantitative metrics of the reconstructions with magnification mismatch for background compensation (r=$16.8~mm$)}
    \renewcommand{\arraystretch}{1.}
    \begin{center}
        \resizebox{0.7\linewidth}{!}{
            \begin{tabular}{lccccccccc}
                \toprule
                Mag. Error                & \percent{-20} & \percent{-15} & \percent{-10} & \percent{-5} & aligned    & \percent{5} & \percent{10} & \percent{15} & \percent{20} \\
                \midrule
                SSIM                      & 0.9828        & 0.9907        & 0.9961        & 0.9991       & \bf0.9999  & 0.9980      & 0.9933       & 0.9866       & 0.9785       \\
                PSNR                      & 26.3718       & 29.2189       & 33.1242       & 39.3940      & \bf49.0959 & 36.5190     & 31.3849      & 28.5055      & 26.5721      \\
                MSE$\times10^{-4}$        & 4.3823        & 2.2750        & 0.9257        & 0.2185       & \bf0.0234  & 0.4236      & 1.3816       & 2.6812       & 4.1848       \\
                RMSRE                     & 0.1043        & 0.0751        & 0.0479        & 0.0232       & \bf0.0076  & 0.0323      & 0.0584       & 0.0813       & 0.1016       \\
                \hline
                SSIM (Crt.)               & 0.9996        & 0.9998        & 0.9999        & 0.9999       & \bf0.9999  & 0.9999      & 0.9998       & 0.9997       & 0.9996       \\
                PSNR (Crt.)               & 41.9195       & 44.8545       & 47.0394       & 48.2090      & \bf48.0992 & 46.9567     & 45.1736      & 43.6877      & 42.4808      \\
                MSE$\times10^{-4}$ (Crt.) & 0.1222        & 0.0621        & 0.0376        & 0.0287       & \bf0.0294  & 0.0383      & 0.0577       & 0.0813       & 0.1074       \\
                RMSRE (Crt.)              & 0.0167        & 0.0122        & 0.0097        & 0.0084       & \bf0.0084  & 0.0095      & 0.0117       & 0.0138       & 0.0159       \\
                \bottomrule
            \end{tabular}
        }
    \end{center}
\end{table*}

\section{Discussions and Conclusion}
In this study, we have proposed a novel CMCT system which incorporates a micro-focus source, PCD, robotic arms and advanced imaging algorithms into a synergistic companion of a conventional CT scanner. The designed HR local scan protocol not only improves the dose efficiency but also reduces the area of detectors. The cost of PCDs currently remains high due to the complex manufacturing techniques, and the situation is likely to continue in the near future. Hence, this hybrid system for interior tomography can reduce the system cost and radiation dose without compromising the required performance. In the local/interior scan, the advanced robotic arms allow the free selection of a VOI, which is preferable compared to the traditional rotating gantry. In addition, the mobility of the robotic system enables surgeons to take projections from any view angle without moving the patient, which can be extremely helpful in many applications, such as high-quality evaluation in emergencies and real-time feedback in surgeries. It is worth mentioning that besides the exemplary application in inner ear imaging, the system can also work for other clinical imaging tasks that demand high resolution in a VOI/ROI, such as tumor examination in breast, nodule characterization in lung, bone quality analysis, and plaque imaging in the heart and the neck.

Besides the hybrid design as one imaging system shown in Fig~\ref{fig:Schematic}, the robotic micro-CT scanner can also be separately used. The reconstruction results from a traditional CT scanner can be used as the prior knowledge following the same procedure described in Section~\ref{sec:reconSTG}. However, this may impose extra work in registration due to different positions of the patient in the local scan and an earlier global scan. Luckily, as analyzed in Section~\ref{sec:Misalign}, the proposed reconstruction method has a good tolerance to geometric misalignment. Although the reconstruction is relatively sensitive to the magnification mismatch, the resultant attenuation shifting can be addressed with the proposed bias correction method. In addition, with the rapid development of face recognition technology \cite{parkhi2015deep}, the human facial surface measurement techniques become mature with high accuracy in real-time \cite{ma2009validation,brahm2017fast, bakirman2017comparison,knoops2017comparison}. The registration between surfaces have been studied for many years. High-quality toolboxes, such as 3D slicer\cite{fedorov20123d}, can be directly used or adapted for our purpose.

\subsection{Projection Deblurring}
Another challenge for CMCT comes from the X-ray source. The intensity of an X-ray source with a micro-focus has usually an insufficient flux to produce a decent signal to noise ratio through a human head during a reasonably short time. To obtain an appropriate contrast, an X-ray source with a slightly larger focal spot may be used to provide enough power. The increased focal spot could generate shadows in the projections and blur structural details. In addition to find the optimal balance between the X-ray intensity and image resolution, a deep-learning based deblurring method can be used to undo the blurring induced by a finite focus size. Although the cone-beam projection with the finite focus spot is no longer a spatially invariant linear system, which is hard to handle with traditional deblurring method, advanced deep learning techniques have the capability of dealing with shift-variant system in deblurring tasks \cite{wang2019deep}. The big data of paired blurred-original projections may be difficult to obtain for training a deblurring network. Alternatively, a forward projection model can be easily built to realistically synthesize the paired data. Then, the network trained with simulated data is fine-tuned with a small amount of paired real projection data. Finally, the trained network can be applied on blurred projections for inference. The feasibility of applying the model trained with synthetic data to the real scenarios has been demonstrated in our previous optical deblurring work \cite{li2020deep}.

\subsection{Image Denoising}
Deep denoising techniques can be used to reduce radiation dose and improve image quality. According to the level of supervision in the network training process, current deep learning methods can be grouped into three categories: supervised learning, weakly-supervised learning, and unsupervised learning.
Supervised learning methods were designed for image denoising and achieved the best performance, such as deep CNNs with residual learning \cite{zhang2017beyond,mao2016image} or with recurrent persistent memory units \cite{tai2017memnet}.
Weakly-supervised learning methods relax the requirement of paired noisy-clean data to unpaired noisy-clean data \cite{chen2018image} or paired noise-to-noise data \cite{lehtinen2018noise2noise}. Using the unpaired noisy-clean data, \cite{chen2018image} proposed GAN-based learning to create pairs of corresponding noisy-clean images as the training data. Recently, \cite{lehtinen2018noise2noise} demonstrated that paired noise-to-noise images are equivalent to the paired noisy-clean images in training a model, achieving a denoising performance competitive with supervised learning methods.
For the applications where even the unpaired noisy-clean or paired noise-noise images are unavailable, unsupervised leaning methods were proposed using only single noisy images for training. Deep image prior \cite{ulyanov2018deep} is a generation process that maps the random noise to a single noisy image, and when they terminate the training process at the right moment the network produces a denoised image. Most recently, Noise2Void \cite{krull2019noise2void} and its variants \cite{krull2019probabilistic,laine2019high} achieved promising results only using individual noisy images in training a network. Basically, a Noise2Void network estimates a blind-spot in an image so that the network learns to map the surrounding pixels to the blind-spot, achieving excellent denoising results.

In our inner ear imaging application, paired noise-clean images can be synthesized via Monte-Carlo simulation, and single real noisy images can be acquired with the proposed CMCT system. The former data type can support supervised training although the noise may not perfectly match the real counterpart, while the latter type of data contains realistic noise and texture. Combining these two types of datasets, we can design a semi-supervised leaning method to learn from the data with and without ground truth labels simultaneously. For example, the model can be trained in the Noise2Void mode first and then fine tuned with the paired noisy-clean data or vice versa. An alternative is to train the model in the Noise2Void and supervised modes simultaneously.

\subsection{Material Decomposition and Beam Harding Correction}
In addition to the high resolution advantage, if the local/interior scan is performed in the multi-channel photon-counting mode, the energy information can be used for spectral analysis; e.g., K-edge imaging, material decomposition, beam hardening correction, and metal artifact reduction. Compared with traditional dual-energy CT, the PCD provides more energy channels and is more informative while the potential spectral distortion issues at high imaging speed can be overcome with deep learning based correction method~\cite{li2020x}. In principle, the direct spectral measurement with the PCD allows better spectral separation than dual-source, fast kVp-switching, and dual-layer detector techniques. Most relevant to this inner ear imaging application is to utilize the X-ray energy dependent attenuation information for beam hardening correction, metal artifact reduction and material decomposition ~\cite{willemink2018photon} so that the effects of the implanted electrodes and micro-environments can be optimally modeled.

\section{Conclusion}
In conclusion, we have proposed a clinical micro-CT (CMCT) system empowered with a number of cutting-edge technologies for regional ultrahigh resolution imaging of a VOI, which is particularly suitable for human inner ear/temporal bone imaging. The background compensation technique has been proposed for fast and accurate local reconstruction with high resolution at a minimized X-ray dose, taking full advantage of prior information from a conventional medical CT scan. The HR feature, attenuation fidelity and robustness to geometric misalignment in the registration between the global and local scans have been demonstrated, establishing the feasibility of CMCT. We also believe advanced deep learning techniques can be leveraged for superior performance, as exemplified by the potential application envisions. Overall, our CMCT system holds a great promise for inner ear imaging as well as other clinical applications.

\section*{Acknowledgment}
We would like to thank for the financial support from National Institutes of Health under award numbers R01CA237267 (NCI) and R01EB026646 (NIBIB), and from the State Scholarship Fund for the China Scholarship Council (CSC No.201906315030).

\appendices
\section*{APPENDIX: ROI reconstructions with mis-alignments}
\begin{figure*}[htbp]
    \centering
    \includegraphics[width=.8\linewidth]{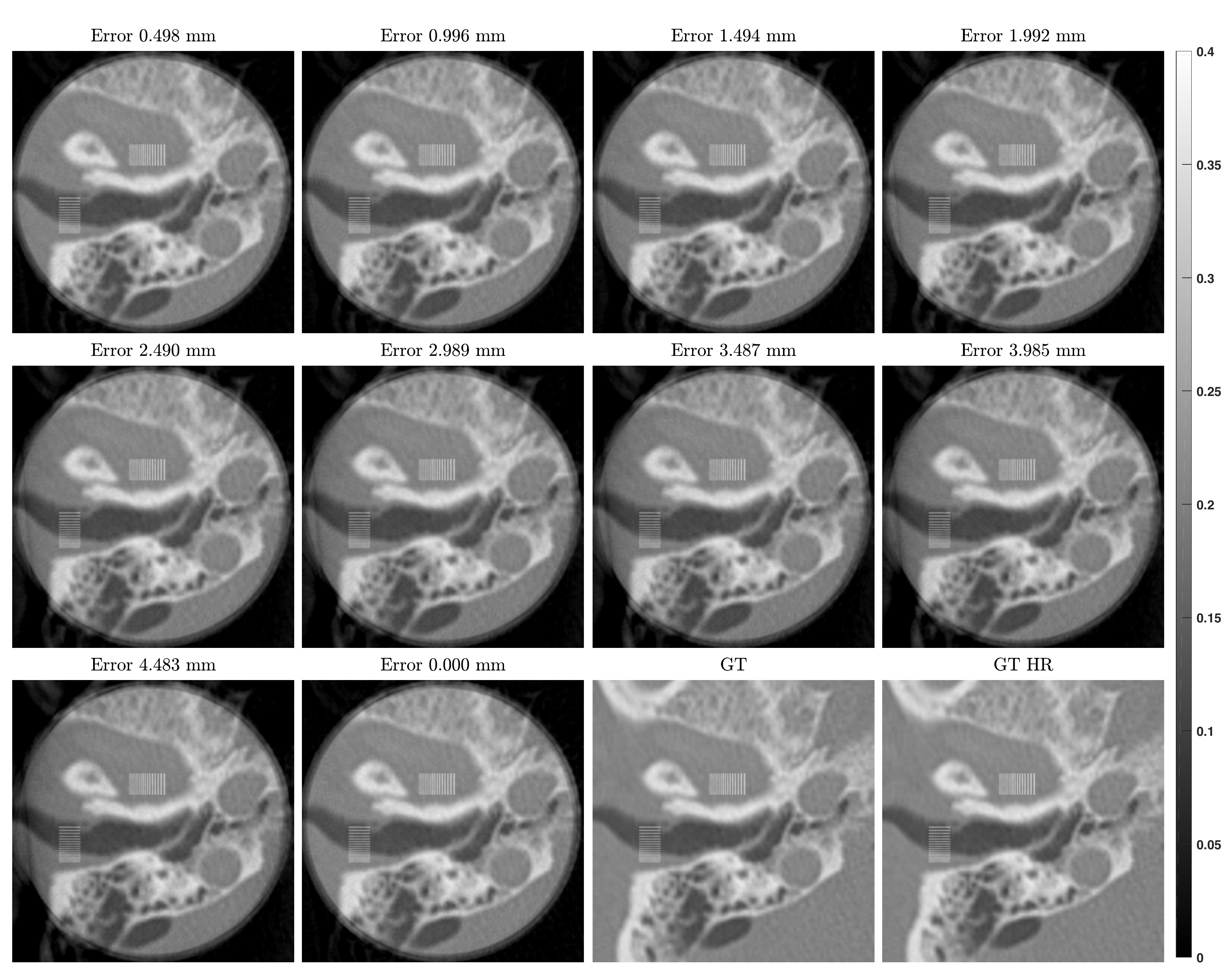}
    \caption{ROI Reconstructions with misalignment of isocenter position in background compensation, together with the ground truths (GT and GT HR), are displayed in the same window (unit of $cm^{-1}$). The misalignment errors are from 0 to \dist{4.483} with an increment of \dist{0.498}.}
    \label{fig:Pimgs}
\end{figure*}

\begin{figure}[htbp]
    \centering
    \includegraphics[width=1.0\linewidth]{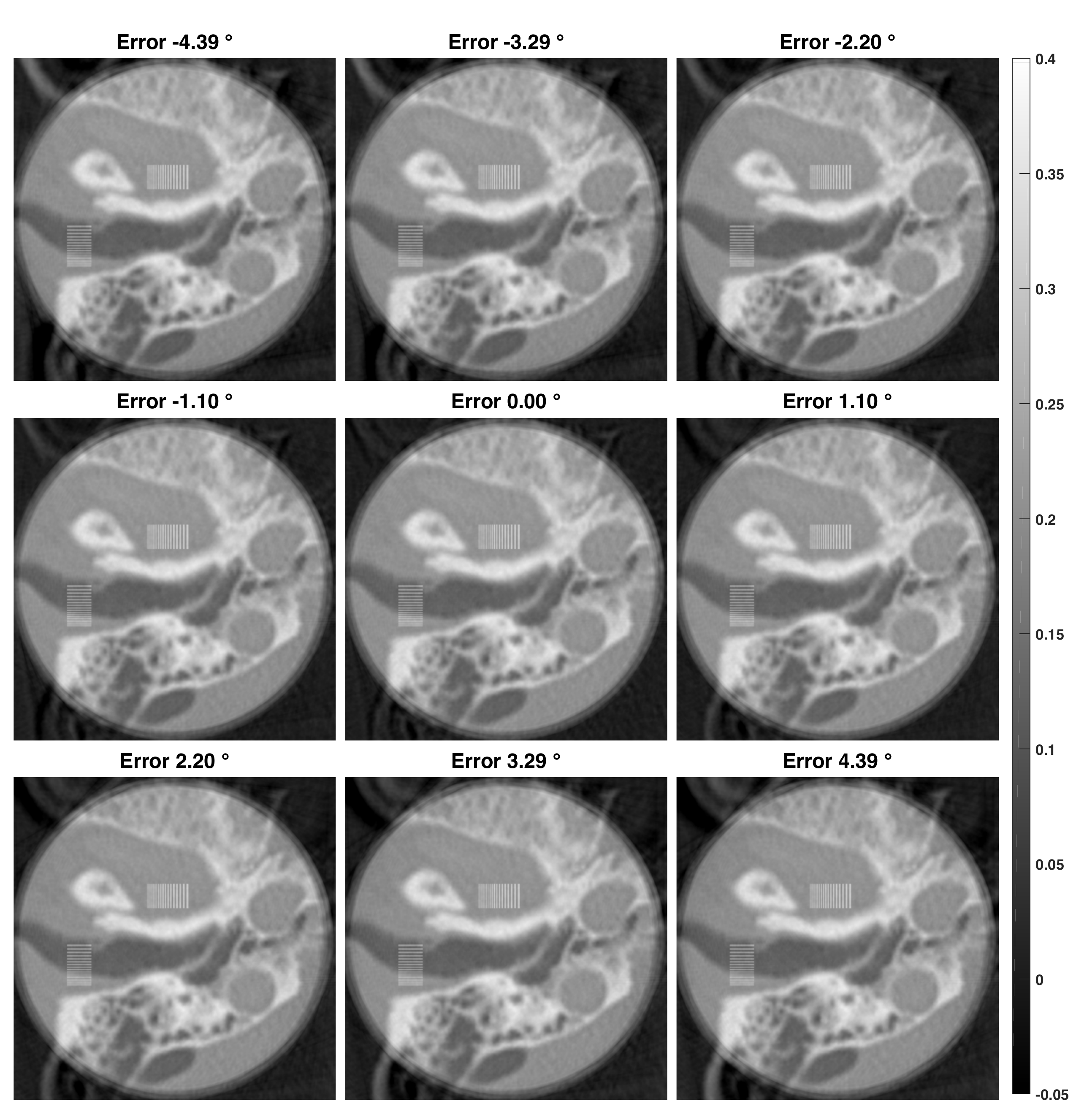}
    \caption{ROI Reconstructions with misalignment of initial view angle in background compensation displayed in the same window (unit: $cm^{-1}$). The misalignment errors are from \degree{-4.39} to \degree{4.39} with an increment of \degree{1.098}.}
    \label{fig:Aimgs}
\end{figure}

\begin{figure}[htbp]
    \centering
    \includegraphics[width=1.0\linewidth]{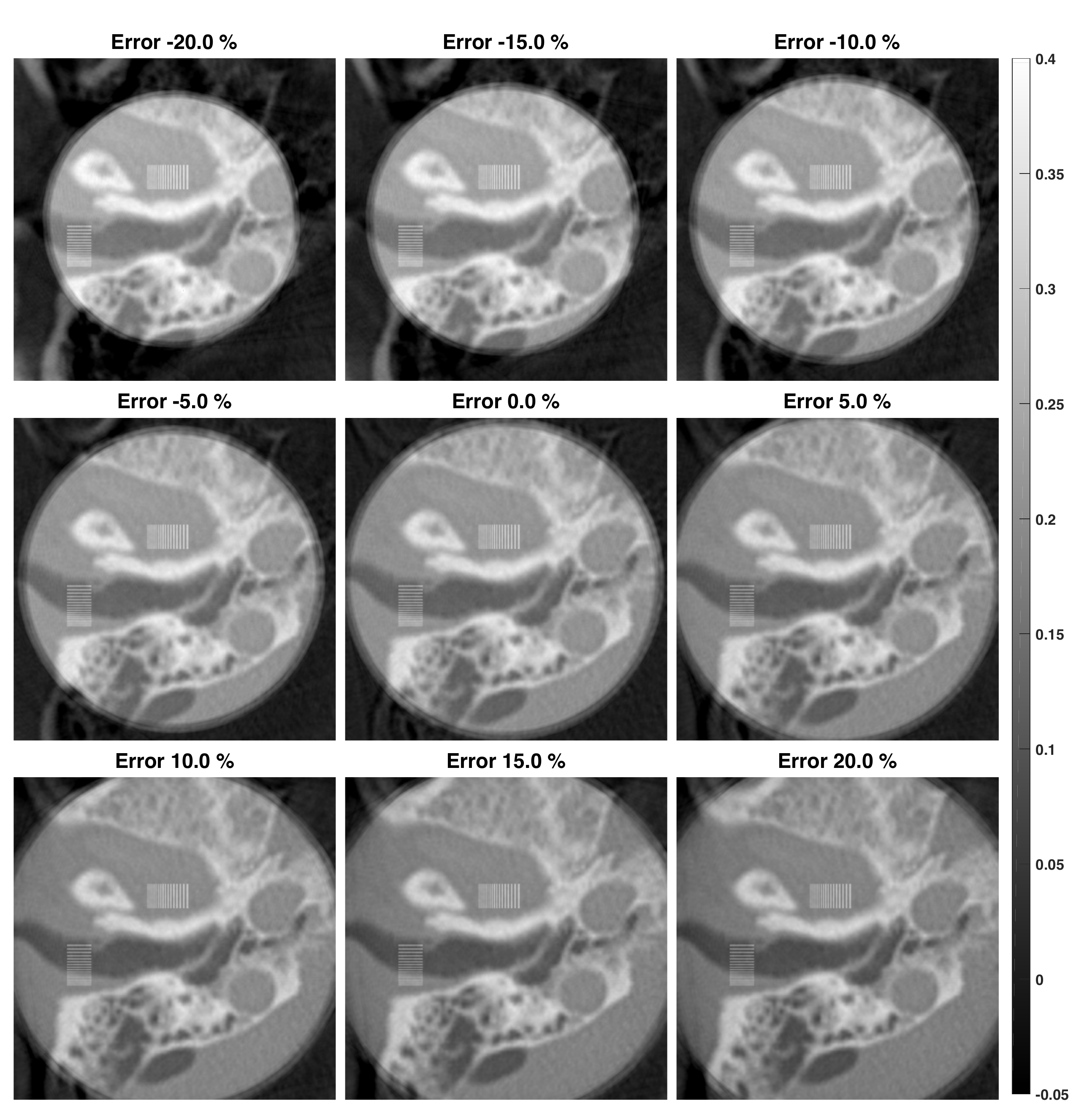}
    \caption{ROI Reconstructions with misalignment of magnification factor in background compensation displayed in the same window (unit: $cm^{-1}$). The misalignment errors are from \percent{-20} to \percent{20} with a step of \percent{5}.}
    \label{fig:Mimgs}
\end{figure}

ROI reconstructions through background compensation with various mis-alignments in terms of isocenter position, initial view angle, and magnification factor, as shown in Fig.~\ref{fig:Pimgs},~\ref{fig:Aimgs} and~\ref{fig:Mimgs}.

\ifCLASSOPTIONcaptionsoff
    \newpage
\fi

\clearpage


\end{document}